# Resonant response in tunable metasurface based on crossed all-dielectric gratings

VLADIMIR V YACHIN[1,2,3,*], VYACHESLAV V KHARDIKOV[1,2], LUDMILA A KOCHETOVA[3], OLEXANDR A MASLOVSKIY[2], KATERYNA L DOMINA[2], AND SERGEY L PROSVIRNIN[1,2]

[1]*Institute of Radio Astronomy, Kharkov 61002, Ukraine*
[2]*Karazin Kharkiv National University, Kharkiv 61022, Ukraine*
[3]*Usikov Institute of Radio Physics and Electronics, Kharkiv 61115, Ukraine*
[*]*Corresponding author: yachin@rian.kharkov.ua*

**We theoretically demonstrate mechanical tuning of the spectral response of a Mie-like resonant dielectric metasurface consisting of crossed all-dielectric strip gratings. We use two array of parallel dielectric strips superimposed on each other to mechanically alter the crossing angle between them thereby changing the position of the maximum of the spectral response in a wide frequency range. The metasurface structure is considered as a doubly-periodic system in an oblique mesh. This mechanically controllable metasurface can be perspective in developing of a flexible polarization-sensitive optical devices.** © 2021 Optical Society of America

http://dx.doi.org/10.1364/ao.XX.XXXXXX

## 1. INTRODUCTION

The metasurface is a two-dimensional analogue of metamaterials with extraordinary properties not found in natural materials [1, 2]. Unlike conventional materials, which have a weak response to external influences and, as a consequence, are able to demonstrate dynamic control of properties in an extremely limited range, metasurfaces open the way to obtain the desired controllability of scattered field properties due to both the cell geometry reconfiguration and material property dynamic changes. As a rule, a change in the geometry of the cell occurs due to mechanical operation. Reversible stretching metasurfaces based on a stretchable substrate resulting in manipulation of transmission spectra by putting a strain along the surface of the polydimethylsiloxane substrate [3, 4]. The use of a micro-electro-mechanical system MEMS extends the functionality of mechanically tunable metasurfaces [5, 6]. Thus, the mechanical reconfiguration of the periodic cell of the metasurface is important. In particular, the effectiveness of this approach to controlling the resonant frequencies without additional external stimuli such as thermal, electrical, magnetic, optical, chemical, or electrochemical and others, discussed, for instance, in [7–9].

With the shortening of the scattering wavelength from microwave to optics, metal periodic structures become less efficient as a result of increasing thermal loss in the metal. One of the most popular research topics for the enhancing efficiency of metasurfaces relates to high-index all-dielectric planar structures with the possibility of expanding their functionality through the use of their reconfigurable properties. Dielectric metasurfaces consist of volumetric elements in which an incident electromagnetic wave excites polarization current resonances associated with magnetic and electric dipole Mie-resonances [10–12]. Recently, metamaterials and metasurfaces with moire configurations, which are regulated by rotation in the plane of superimposed layers, have drawn attention to the extraordinary optical properties of such quasi-periodic structures, including multiband and broadband responses and optical activity [13, 14].

In our letter we continue our research of the mechanical tunable crossing gratings. In [15] we demonstrate for metallic crossing strip gratings moving resonant peak along frequency scale through altering the angle of crossing. This resonance is associated with the length of the rhombus side of the structure cell when the lattice symmetry is broken. In contrast with those resonances when replacing metal strips with a dielectric one we observe volumetric Mie-like resonance that leads to the same effect namely moving resonant peak along frequency scale through altering the angle of strips crossing.

## 2. FORMULATION OF THE PROBLEM

In Fig. 1 we present the scheme of an all-dielectric metasurface formed by the intersection of two identical periodic arrays ($a$ is arrays period) of parallel dielectric strips of rectangular cross section. To simplify the physical analysis of the formation of the response of such metasurface to electromagnetic excitation, substrates on which periodic arrays can be located are excluded from consideration.

The periodic cell of the metasurface under study has the shape of a rhombus with an acute angle $\beta$ and a side $L = L_x = L_y = a/sin(\beta)$ that defines the period of the metasurface along the $Ox_1$ and $Oy_1$ axes . Dielectric strips of thickness $h = 0,05L$ and width $D = a/3$ are considered to be made of non-magnetic material with $\mu = 1$ and with a dielectric constant that corresponds to the dielectric constant of silicon at the terahertz range $\varepsilon = 11.0$ and $tan\delta = 10^{-3}$ (the time dependence is assumed in the form $\exp(-i\omega t)$). The metasurface can be tuned by mechan-ically rotating one of the arrays at an arbitrary angle relative to the other. We assume that the backward array ($0 \leq z \leq h$) contains strips that are oriented along the $Oy$ axis and are fixed, and the front array ($h \leq z \leq 2h$) can be rotated in the $xOy$ plane. It is clear that different metasurfaces will be formed by varying the values of angles $0 \leq \beta \leq 90°$. A plane electromagnetic wave incidents on the metasurface from the upper region $z > 2h$ in the opposite direction to the $Oz$ axis. We will distinguish cases of $x$- and $y$- polarized waves.

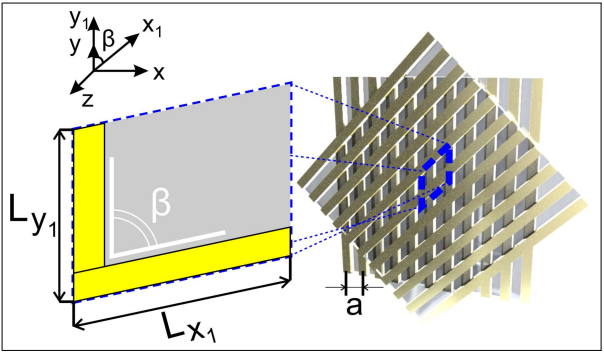

**Fig. 1.** The metasurface with oblique periodic cell. The dotted frame indicates the unit cell of the structure.

## 3. THE NUMERICAL SIMULATIONS

Analysis of the resonant response of the studied structure to electromagnetic excitation was performed on the basis of numerical simulated scattering of a plane wave on a periodic structure, using the well-tested method of integral functionals [16]. The method uses volume integral equations for equivalent electric and magnetic polarization currents induced by the incident wave field in the periodic layer. Integral equations are solved numerically with the help of integral functionals related to the distribution of the polarization current and the technique of Floke-Fourier series double decomposition. At the final stage of the solution, we obtain scattered fields as a superposition of diffraction harmonics in the following form:

$$\mathbf{E}^r(x,y,z) = \sum_{p=-\infty}^{\infty} \sum_{q=-\infty}^{\infty} \mathbf{E}^r_{pq} e^{i(\xi_p x + \gamma_{pq} y - \kappa_{pq} z)}, \quad (1)$$

$$\mathbf{E}^t(x,y,z) = \sum_{p=-\infty}^{\infty} \sum_{q=-\infty}^{\infty} \mathbf{E}^t_{pq} e^{i(\xi_p x + \gamma_{pq} y + \kappa_{pq} z)}, \quad (2)$$

where
$\xi_p = k_x + \frac{2\pi p}{L_{x_1}}$, $\gamma_{pq} = k_y + \frac{2\pi}{\sin\beta}\left(\frac{q}{L_{y_1}} - \frac{p}{L_{x_1}}\cos\beta\right)$, $\kappa_{pq} = \sqrt{k^2 - \xi_p^2 - \gamma_{pq}^2}$. are the wave vector components of a diffracted order p and q, $\mathbf{E}^r_{pq}$, and $\mathbf{E}^t_{pq}$ that are the magnitudes of the electric field of the reflected and transmitted waves.

In Fig. 2 one can see that in the case of the incident of $x$-polarized wave on the structure, the resonance formed in the reflectance spectrum shifts to the high-frequency region with increasing angle $\beta$. This resonance for the case of $y$-polarization has much smaller amplitude. The conversion into cross-polarization electromagnetic waves is almost the same, for both polarised waves which allows us to speak about the polarization independence of the polarization-transformation properties of such structures. Note that the polarization-transformation properties of such structures also posses a resonant character and for the reflective metasurfaces, in which the strips array can be placed on a metal substrate, these patterns are preserved.

For any value of the angle $\beta$ except for $\beta = 0$ and $\pi/2$, the investigated structure manifests properties inherent to objects with a volume chirality. The relevant properties appear as a result of electromagnetic coupling between closely placed gratings of a double-layered structure [17]. Obviously, the sign of chirality is defined by the sign of the angle $\beta$. A circular dichroism and an optical activities of the metasurfase may be observed as consequence of chirality. Both effects are controllable in the proposed metasurface.

In terms of circular polarised waves, a transmission matrix may be presented by using calculated transmission coefficients of linear polarised waves in accordance with the following expression [18].

$$\begin{pmatrix} t_{++} & t_{+-} \\ t_{-+} & t_{--} \end{pmatrix} = \quad (3)$$

$$= \frac{1}{2}\begin{pmatrix} t_{xx} + t_{yy} + i(t_{xy} - t_{yx}) & t_{xx} - t_{yy} - i(t_{xy} + t_{yx}) \\ t_{xx} - t_{yy} + i(t_{xy} + t_{yx}) & t_{xx} + t_{yy} - i(t_{xy} - t_{yx}) \end{pmatrix}$$

where indexes $+$ and $-$ denote right-handed and left-handed circularly polarised waves respectively.

The circular dichroism can be calculated by expression

$$D = |t_{++}|^2 - |t_{--}|^2. \quad (4)$$

The polarisation azimuth angle can be found by formula

$$\Phi = -\frac{1}{2}[\arg(t_{++}) - \arg(t_{--})]. \quad (5)$$

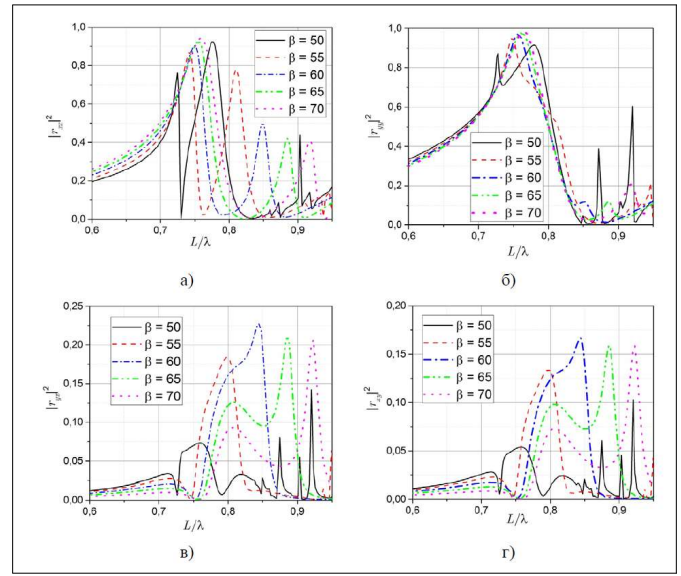

**Fig. 2.** The normalised frequency dependences of the reflectance coefficients ($|r_{ji}|^2$) of the co- (a) $x$- and (b) $y$- polarized waves and cross- conversion coefficients for the cases of normal incidence on the structure of the (c) $x$- and (d) $y$- polarized waves with different angle $\beta$. Here the indices $i$ and $j$ take on the values $x, y$.

If shown in Fig. 2 the spectral responses compared with those for the metasurface with a square periodic cell (Fig. 3) we can assume that in this case we are dealing with the removal of the degeneracy of resonances inherent in a square cell due to its transformation into a rhombic one with splitting of one resonance into two. In this case, we can also assume that the peculiarities of the reflection behavior of the polarized wave are due to the fact that the electric field vector is not parallel to the dielectric strips, and therefore the reflection pattern is due to interference of natural oscillations in two directions corresponded the directions of crossed strips. In order to verify these assumptions, we analyze the distributions of fields in the metasurface that correspond to the absorption maxima in the structure.

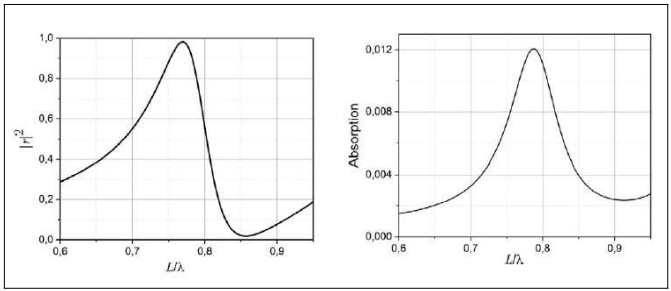

**Fig. 3.** The spectral responses of (a) the reflection coefficients (b) the absorption coefficients for the normal incidence of an arbitrary polarized electromagnetic wave on a metasurface with a square shape of an elementary cell.

Fig. 4 shows the distributions of electric fields in a metasurface with a square periodic cell ($\beta = 90°$) at a frequency corresponded to the absorption maximum in Fig. 3 (b). We see that the electric dipole field distribution is formed in the overlap region of the dielectric strips, which lies in the plane of the metasurface. The orientation of this electric dipole coincides with the direction of the electric field vector in the wave incident on the metasurface. It should be noted that in this case the spectral response of the reflection and absorption coefficients does not depend on the polarization of the wave due to the symmetry of the problem.

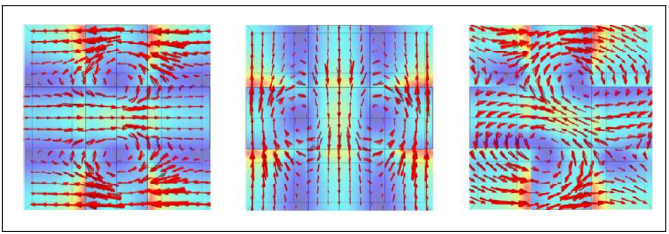

**Fig. 4.** Distributions of the amplitude (color map) and the vectors of the electric field in the plane $z = h$ at the frequency of maximum absorption of the metasurface with a square cell ($L/\lambda = 0.79$) (a) for the case of $x$-polarized wave, (b) the same for the $y$- polarized wave, and (c) when vector of the electric field **E** forms 30° angle between the $x$- direction.

Fig. 5 shows the distributions of the vectors and the amplitudes of the electric field in the periodic cell of a metasurface with a rhombic periodic cell ($\beta = 50°$) at the frequencies of the observed absorption resonances, there are 2 in this case, for three different polarizations of the incident wave.

It should be noted that, as in the previous case, the absorption maxima for all 3 polarizations of the incident wave coincide,

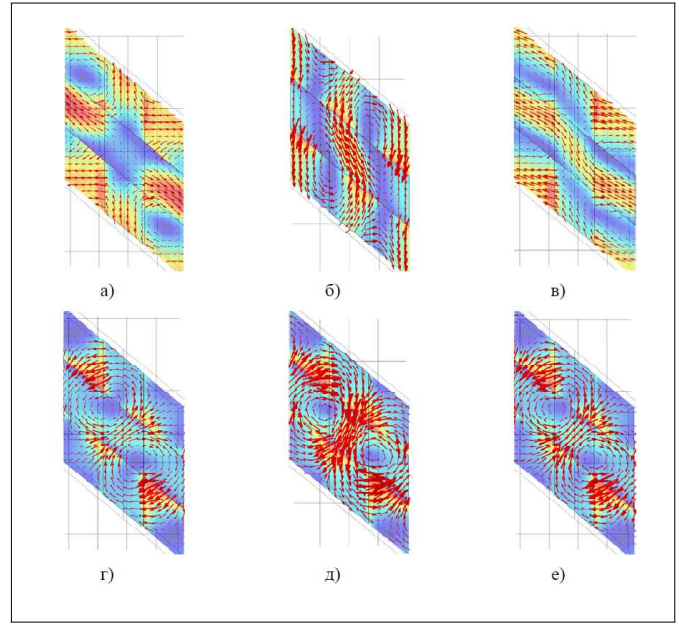

**Fig. 5.** Distributions of the amplitudes (color map) and vectors (arrows) of the electric field in the unit cell at the frequencies of the absorption maxima of the metasurface (a, b, c correspond to $L/\lambda = 0.78$, d, e, f - $L/\lambda = 0.73$) ) with a rhombic cell ( $\beta = 50°$) for the case (a, d) $x$- polarized, (b, e) $y$- polarized waves and (c, f) the vector of the electric field **E** directed along the $Ox_1$ axis forms the angle $\beta = 50°$ between the $Oy$ axis.

and, in turn, suggests that the resonant response of the metasurface is associated with the formation of intrinsic resonances of metasurface elements. In addition, it is easy to see that all resonances are associated with the formation at the intersection of dielectric strips that are electric dipoles lying in the plane of the metasurface, and posses the same nature as the resonance observed in the case of a metasurface with a square periodic cell. Analysis of the field distributions allows us to see that the high-frequency resonance is associated with the excitation of the dipole along the longer diagonal of the rhombus, while the low-frequency resonance is associated with the excitation of the dipole along the smaller diagonal. That is, we really observe the removal of the degeneracy of the two resonances due to the deviation of the angle of intersection of the arrays from $\beta = 90°$.

The distribution pattern of the field in the case of an $x$- polarized wave is explained by the fact that in this case we observe the superposition of electromagnetic fields which occur when excited resonances by the incident waves with an electric field vector parallel to the strips of one of the two arrays. It is easy to see that for both types of resonance these distributions have antiphase dipoles in the region of strip overlap, which leads to the observed result.

In summary, in this Letter, we present for the first time achieving mechanical tuning of the resonant response of an all dielectric metasurface at the terahertz range by removing the degeneracy of the electric dipole resonance of crossed arrays with silicon strips through their rotation relative to each other. Mechanical frequency tuning of the resonant peak position occurs in a fairly wide frequency range. This kind of metasurfaces can be a key component in planar configurable optical devices.

**Acknowledgments.** Authors are grateful to the National Research

Fund of Ukraine for support of this work partly by the project grant 2020.02/0218.

**Disclosures.** The authors declare no conflicts of interest.

## FULL REFERENCES